\documentclass[preprint,aps,showpacs]{revtex4}
\usepackage{graphicx}
\usepackage{subfigure}
\usepackage{dcolumn}

\usepackage[T1]{fontenc}
\usepackage{bm}

\usepackage{amsmath}
\usepackage{amsfonts}
\usepackage{float}
\usepackage{latexsym}
\usepackage{amssymb}
\usepackage{epsfig}
\usepackage{color}
\usepackage{parskip}
\usepackage{tabularx}

\usepackage{times}
\begin{document}
\title{Phase-sensitive amplification via coherent population oscillations in metastable helium at room temperature}
\author{J. Lugani$^1$}
\email{jasleen.lugani@u-psud.fr}
\author{C. Banerjee$^1$}
\author{M-A. Maynard$^1$}
\author{P. Neveu$^1$}
\author{W. Xie$^{1,2}$}
\author{R. Ghosh$^3$}
\author{F. Bretenaker$^1$}
\author{F. Goldfarb$^1$}
\affiliation{$^1$Laboratoire Aime Cotton, CNRS - Universite Paris Sud 11 - ENS Cachan - Universite Paris-Saclay, 91405 Orsay Cedex, France\\
$^2$State key Laboratory of Advanced Optical Communication Systems and Networks, Shanghai Jiao Tong University, Shanghai, 200240, China\\
$^3$Shiv Nadar University Gautam Buddha Nagar, UP 201314, India}
\begin{abstract}
\normalsize

In this letter, we report our experimental results on phase-sensitive amplification (PSA) in non-degenerate signal-idler configuration using ultra-narrow coherent population oscillations in metastable helium at room temperature. We achieved a high PSA gain of nearly 7 with a bandwidth of 200 kHz, by using the system at resonance in a single-pass scheme. Further, the measured minimum gain is close to the ideal value, showing that we have a pure PSA. This is also confirmed from our phase-to-phase transfer curves measurements,  illustrating that we have a nearly perfect squeezer, which is interesting for a variety of applications.
\end{abstract}

\maketitle
Over the last few years, phase-sensitive amplification (PSA) has been a subject of wide research in a variety of fields due to its unique noise properties. It enables amplification of a weak signal without adding any extra noise, i.e. without degrading its signal to noise ratio~\cite{Caves}, and thus finds applications in metrology~\cite{metrolgy}, imaging industry~\cite{imaging} and telecommunication~\cite{PA2011}. Further, this noiseless amplification is associated with the generation of squeezed states of light, which makes this parametric process very interesting for quantum optics and quantum information experiments~\cite{Agarwal}. PSA has been successfully achieved in nonlinear crystals and waveguides~\cite{levenson} through three-wave mixing ($\chi^{(2)}$ process), and in fibers~\cite{TongIEEE2012} and alkali vapors such as rubidium~\cite{PLprl,PLOE1} through four-wave mixing (FWM) ($\chi^{(3)}$ process). Although a very large quantum noise reduction has been achieved using crystals~\cite{Hemmer}, it is difficult to directly couple the down-converted photons with atomic systems because of their frequency and bandwidth, while it would be useful for many applications in quantum information such as realization of atomic memories, processing of atomic qubits through quantum light, entanglement swapping, etc. Realizing PSA in the same atomic system is thus interesting as the generated non-classical light is automatically within the bandwidth of the atomic transition, spectrally narrow, and can moreover be spatially multimode~\cite{PLOE1, JingPRL, Novikova16}. 
 
Motivated by these works, we report our results on PSA in metastable helium (He$^{4}$) at room temperature, through coherent population oscillation (CPO) assisted FWM processes in a $\Lambda$ scheme at resonance. In other atomic systems (e.g. alkali vapors), the FWM process relies on the coherence between the two ground states of the $\Lambda$ system~\cite{PLprl,PLOE1,JingPRL}. In comparison to this, we use a particular kind of CPO resonance, which involves the dynamics of the population difference of the atoms in the two ground states~\cite{thomas12} to enhance the non-linearity of the system. It enables us to achieve comparable PSA gains with approximately 2 $\times$ 10$^{11}$ cm$^{-3}$ of atomic density, which is at least two orders of magnitude less than rubidium~\cite{PLOL1}. Further, $^4$He has other favorable properties such as absence of nuclear spin resulting in a simplified energy level structure without any hyperfine levels. This has an important consequence as it eliminates unwanted FWM processes, which usually arise due to transitions in the nearby hyperfine levels and add extra noise and degrade squeezing~\cite{PLOE1}. Thus, using this simple system, we expect to achieve high PSA gains, close to resonance, within the Doppler width and implement a perfect squeezer.
\begin{figure}[htbp]
\centering
\includegraphics[width=12cm]{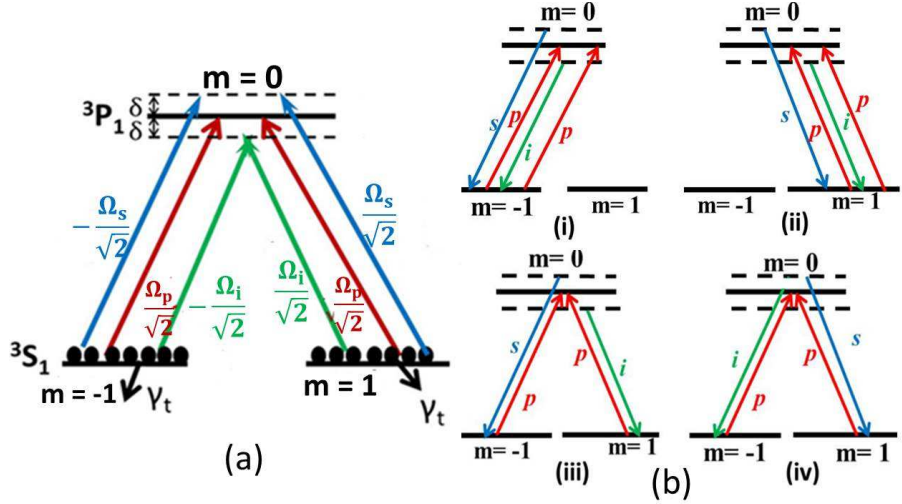}
\caption{(a) (Color online) Schematics of the $\Lambda$ structure in helium, both arms are excited by all the beams: pump ($\Omega_p$, red), signal (($\Omega_s$), blue) and idler ($\Omega_i$, green) (b) Different FWM processes possible in the system}
\label{fw}
\end{figure}
	
	The experiment is based on the $2^3\mathrm{S}_1 \rightarrow 2^3\mathrm{P}_1$ (D1) transition of He$^{4}$, and the $\Lambda$ system is constituted by two transitions corresponding to $\sigma^+$ and $\sigma^-$ polarizations (see Fig. \ref{fw}a). We excite this system with linearly polarized pump, signal, and idler beams with co-polarized signal and idler beams orthogonal to the pump beam. 
A FWM process takes place when two pump photons are annihilated and a signal and an idler photon are generated, or vice versa: due to the chosen polarizations of the beams, there are four FWM channels possible in this $\Lambda$ system, each conserving energy and momentum (both linear and angular), as shown in fig. \ref{fw}b. In process (i), two $\sigma^+$ pump (p) photons are absorbed in the left arm and a $\sigma^+$ signal ($s$) and a $\sigma^+$ idler ($i$) photon are emitted in the same arm. Likewise, in process (ii) all photons are $\sigma^-$ and are on the right arm. These two processes are based on CPO in the coupled open system~\cite{thomas12}. In process (iii) (and (iv)) two pump photons, a $\sigma^+$ and a $\sigma^-$, are absorbed and a $\sigma^+$ ($\sigma^-$) signal and a $\sigma^-$ ($\sigma^+$) idler photon are emitted. Processes (iii) and (iv) involve the excitation of Raman coherence between the two ground states. For the present work, we have not performed any quantum measurements and thus, the system can be modeled using classical approach. But in view of the future applications in the quantum domain, we adopt here a quantum-mechanical approach and the qualitative behavior of the PSA can be explained using a similar formalism as in \cite{levenson,Jingth}. Considering the pump as a strong classical field  and signal and idler beams as quantum mechanical operators, the FWM interaction Hamiltonian can be written as~\cite{Jingth} 
\begin{equation}
\hat{H}=i \hbar \zeta e^{2 \phi_{p_{in}}} \hat{a_s} \hat{a_i}+ h.c,
\label{hami}
\end{equation}
where $\hat{a_s} (\hat{a_i})$ is the annihilation operator corresponding to the signal (idler), $\zeta$ is the strength of the FWM process, proportional to the third order susceptibility and the intensity of the pump beam and $\phi_{p_{in}}$ is the input phase of the pump. In the Heisenberg picture, using the Hamiltonian (Eq. \ref{hami}), the rate equation for $\hat{a_s}$ is given as d$\hat{a_s} (t)/$d$t=i/\hbar [\hat{H},\hat{a_s}(t)]$, from which its time evolution can be evaluated as:
\begin{eqnarray}
\hat{a_s}(t)= \cosh (\zeta t)\hat{a}_{s0} + e^{i 2\phi_{p_{in}}} \sinh (\zeta t) \hat{a}_{i0}^\dagger,\nonumber\\
\hat{a_i}^\dagger (t)= e^{-i 2\phi_{p_{in}}} \sinh (\zeta t) \hat{a}_{s0} + \cosh (\zeta t) \hat{a}_{i0}^\dagger.
\label{ops}
\end{eqnarray}
Neglecting pump depletion, PSA gain for the signal is then computed by calculating the ratio between the average number of signal photons at the output $\left\langle \hat{a_s}^\dagger (t) \hat{a_s} (t) \right\rangle$ and the input $\left\langle \hat{a}_{s0}^\dagger \hat{a}_{s0} \right\rangle$ and is found to be
\begin{equation}
G_{PSA}=2g - 1 + 2 \sqrt{g(g-1)} \cos{(\Phi)},
\label{eqgain}
\end{equation}
where $g=(\cosh(\zeta t))^2$ and $\Phi=2\phi_{p_{in}}-\phi_{s_{in}}-\phi_{i_{in}}$ is the relative phase between the three beams: pump, signal and idler. Thus, from the above equation (Eq.(\ref{eqgain})), we see that the gain is maximum ($G_{max}$) when $\Phi=0$ and minimum ($G_{min}$) when $\Phi=\pi$ and that for an ideal PSA, $G_{min}=1/G_{max}$~\cite{levenson}.  
\begin{figure}[htbp]
\centering
\includegraphics[width=12cm]{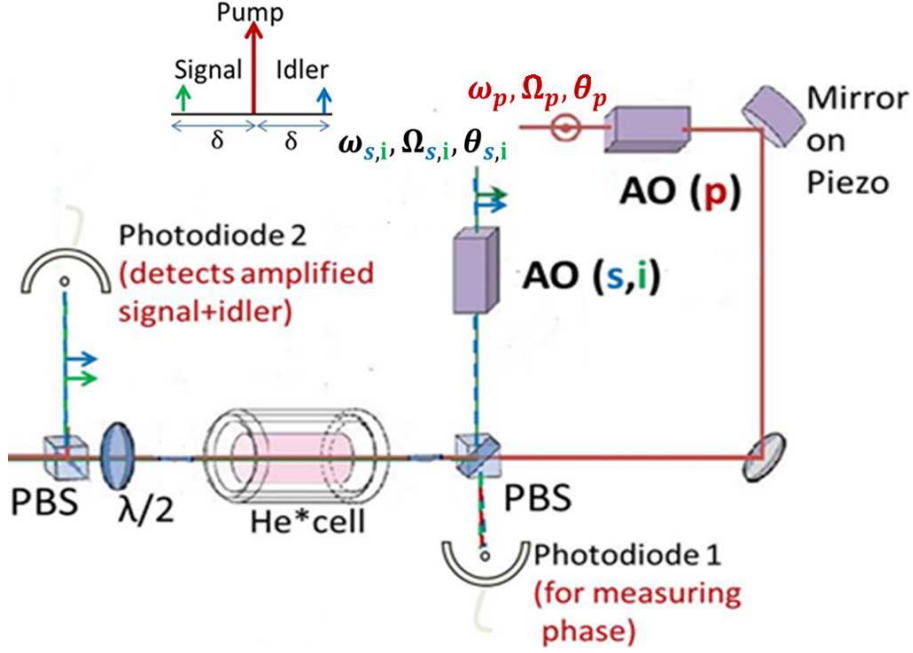}
\caption{(Color online) Experimental setup: Pump, signal and idler are derived from the same laser with frequencies and amplitudes controlled by two acousto optic (AO) modulators. Signal and idler are non-degenerate but identically linearly polarized (orthogonal to the pump polarization) and follow the same optical path. A polarizing beam splitter (PBS) recombines the beams before the cell. The piezo-actuator in the pump path enables to scan the phase. The photodiode 1 is used to  measure the input relative phase between pump and signal/idler before the cell and the amplified output is detected after the cell by the photodiode 2}
\label{exp}
\end{figure}

\begin{figure}[htbp]
\centering
\includegraphics[width=14cm]{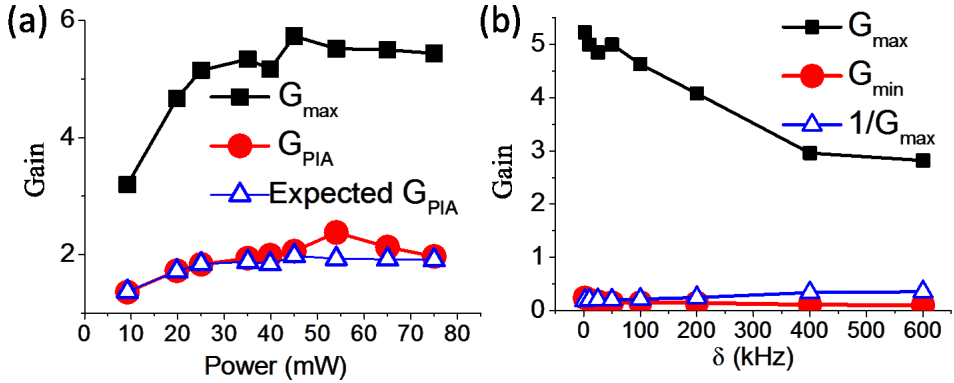}
\caption{(Color online) (a) Variation of maximum PSA gain ($G_{max}$, black squares) and PIA gain ($G_{PIA}$, red circles) and expected PIA gain from $G_{max}$ (blue triangles) as a function of pump power (b) PSA Gain spectrum: Variation of $G_{max}$ (black squares) and $G_{min}$ (red circles) and 1/$G_{max}$ (blue triangles) as a function of pump-signal detuning ($\delta$)}.   
\label{piaf}
\end{figure}

 Our experimental set up is shown in fig. \ref{exp}. The helium cell is 6 cm long, filled with 1 Torr of He$^4$ and is at room temperature. It is placed in a three-layer $\mu$-metal shield to remove magnetic field gradients. Helium atoms are excited to the metastable state by an RF discharge at 27 MHz. The Doppler width corresponding to the optical transition (D1) is around 0.9 GHz (half width at half maximum). For the non-degenerate signal-idler PSA configuration, signal and idler photons have a frequency separation of $2\delta$ and they are symmetrically located on either side of the pump frequency ($\omega_p$) as shown in fig. \ref{exp}. Both beams are derived from the same laser at 1.083 $\mu$m and have nearly the same diameter of about 2 mm. They are controlled in frequencies and amplitudes by two acousto-optic modulators (AO) and recombined using a polarizing beam splitter (PBS). The pump power can be varied from 5 mW to 80 mW using a tapered amplifier while the signal and idler have equal powers at the input of the cell, around 30 $\mu$W. The input relative phase ($\Phi$) between the pump, signal and idler is scanned using a piezo actuator attached to a mirror in the pump path and is measured using the beatnote detected by the photodiode 1 before the cell (see fig. \ref{exp}). After the cell, polarization optics allows the detection of mainly the amplified signal and idler along with a small amount of coupling. Using this residual coupling as the local oscillator, we perform heterodyne detection and measure the output relative phase of the amplified signal/idler with respect to the pump. Thus, at photodiode 2, we detect the beating between the three beams: signal, idler and residual pump, reading: 
\begin{eqnarray}
I=G_s I_{s_{in}}+G_i I_{i_{in}}+2\sqrt{G_s I_{s_{in}} G_i I_{i_{in}}} \cos{(2\delta t+\Delta \phi_{si})}+ I_{p}  \nonumber\\
+2 \sqrt{I_p} (\sqrt{G_s I_{s_{in}}} \cos(\delta t + \Delta \phi_{sp_{out}})+\sqrt{G_i I_{i_{in}}}\cos(\delta t + \Delta \phi_{ip_{out}})),
\label{eq1}
\end{eqnarray}
where $G_{s(i)}$ is defined to be the signal (idler) gain, as the ratio of the output signal (idler) intensity to the input signal (idler) intensity. $I_{s_{in}}$ and $I_{i_{in}}$ correspond to the input signal and idler intensities, respectively, and $I_p$ is the residual pump intensity. For PSA operation, we send equal intensities of signal and idler with the same phase, i.e. $I_{s_{in}}=I_{i_{in}}$ (and $G_s=G_i=G$). We checked that the relative phase between signal and idler is still 0 at the output when $\delta$ is small enough (< 25 kHz), i.e. $\Delta \phi_{si}=0$. This also results in the same phase for the pump-signal and pump-idler beatnote at the output, i.e. $ \Delta \phi_{sp_{out}}=\Delta \phi_{ip_{out}}  \equiv \Delta \phi_{out}$. Thus Eq.(\ref{eq1}) reduces to
\begin{equation} 
I=2 G I_{s_{in}}+2 G I_{s_{in}} \cos{(2\delta t)}+ I_{p}+ 4 \sqrt{I_p G I_{s_{in}}} \cos{(\delta t)}  \cos{(\Delta \phi_{out})}.
\label{eqI}
\end{equation}
In order to evaluate PSA gain $G$, we perform Fourier transform of the data, which gives us peaks at $\delta$ and $2\delta$ frequencies. The gain is then calculated by computing the ratio of the amplitudes of the peak at $2\delta$ frequency for the cell-on and cell-off conditions.   
\begin{figure}[htbp]
\centering
\includegraphics[width=14cm]{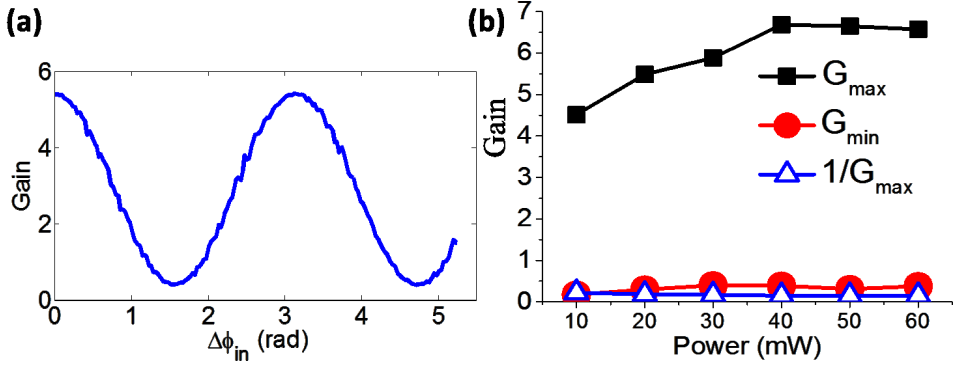}
\caption{(Color online) (a) Variation of PSA gain as a function of input relative phase between pump and signal ($\Delta \phi_{in}$) for a pump power of 30 mW (b) Variation of maximum PSA gain ($G_{max}$, black squares) and minimum PSA gain ($G_{min}$, red circles) as a function of pump power. 1/$G_{max}$ ( blue triangles) corresponds to the ideal value for $G_{min}$}
\label{psf}
\end{figure}
 For PSA, as the relative phase ($\Phi$) between the pump, signal and idler at the input is scanned, the signal successively undergoes amplification ($G>1$) and deamplification ($G<1$). Note that in the present case, by our definition, $\Phi=2\phi_{p_{in}}-\phi_{s_{in}}-\phi_{i_{in}}=2\phi_{p_{in}}-2\phi_{s_{in}}=2\Delta \phi_{in}$, where $\Delta \phi_{in}$ is the relative phase between pump and signal at the input. In the experiment, the piezo actuator attached to the mirror in the pump path scans the relative phase and we study the variation of the gain as shown in fig. \ref{psf}a, which is as expected theoretically (Eq.(\ref{eqgain})). The maximum obtainable PSA gain depends on the input pump power, the overlap between the spatial modes of the beams and the optical detuning. We have studied the variation of the maximum and minimum gains ($G_{max}$ and $G_{min}$, respectively) as a function of pump power as shown in fig. \ref{psf}b. A maximum gain of around 7 can be achieved for 40 mW of pump power and a pump-signal detuning ($\delta$) of 2 kHz. With better alignment and larger optical thickness, we may achieve even larger gains, for example through a multi-pass scheme. From fig. \ref{psf}b, it is visible that the measured $G_{min}$ is close to the ideal value (=1/$G_{max}$) for a wide range of pump powers. Further, we have also performed phase insensitive amplification (PIA) in our scheme, by sending only the signal (and no idler) at the input of the cell. In this case, the idler is generated at a frequency ($\omega_p-\delta$) from the vacuum fluctuations and the signal is amplified in the process. Under these conditions, PIA gain can be similarly found using Eq.(\ref{ops}) with no idler at the input, i.e. by substituting $\hat{a}_{i0}=0$. The resulting PIA gain is then given as $G_{PIA}$=g=$(\cosh(\zeta t))^2$ and from Eq.(\ref{eqgain}), the relationship between the maximum PSA and PIA gain is: $G_{max}=2G_{PIA}-1+2\sqrt{G_{PIA}(G_{PIA}-1)}$. In the experiment, the PIA gain is measured for different pump powers and is found to be close to its ideal value obtained from the corresponding $G_{max}$ as shown in fig.\ref{piaf}a. We have also investigated the variation of gain with the pump-signal detuning, $\delta$ (fig. \ref{piaf}b, for pump power=30 mW). We define the PSA bandwidth as the maximum value of $\delta$ separation for which $G_{min}$ is very close to its ideal value (1/$G_{max}$). As shown in fig. \ref{piaf}b, the system has a large gain bandwidth of more than 200 kHz. It is to be noted that this agrees well with the bandwidth of the CPO resonance at the corresponding pump power~\cite{thomas12}. For larger $\delta$ separation, the signal frequency goes out of the transparency window and gets absorbed and thus both maximum and minimum gains tend to drop as shown in fig. \ref{piaf}b. The gain bandwidth can be increased by improving the spatial modes of the beams and the alignment at the input.
We must stress here that unlike other atomic systems, our scheme does not suffer from 'unwanted' FWM processes~\cite{PLprl,PLOE1}. This is evidenced from (a) the absence of any additional peaks at any undesired frequency in the Fourier transform of the beatnote pattern detected at photodiode 2, (b) the fact that, in our system, $G_{min}$ $\approx$ 1/$G_{max}$ for a wide range of experimental parameters.
\begin{figure}[htbp]
\centering
\includegraphics[width=12cm]{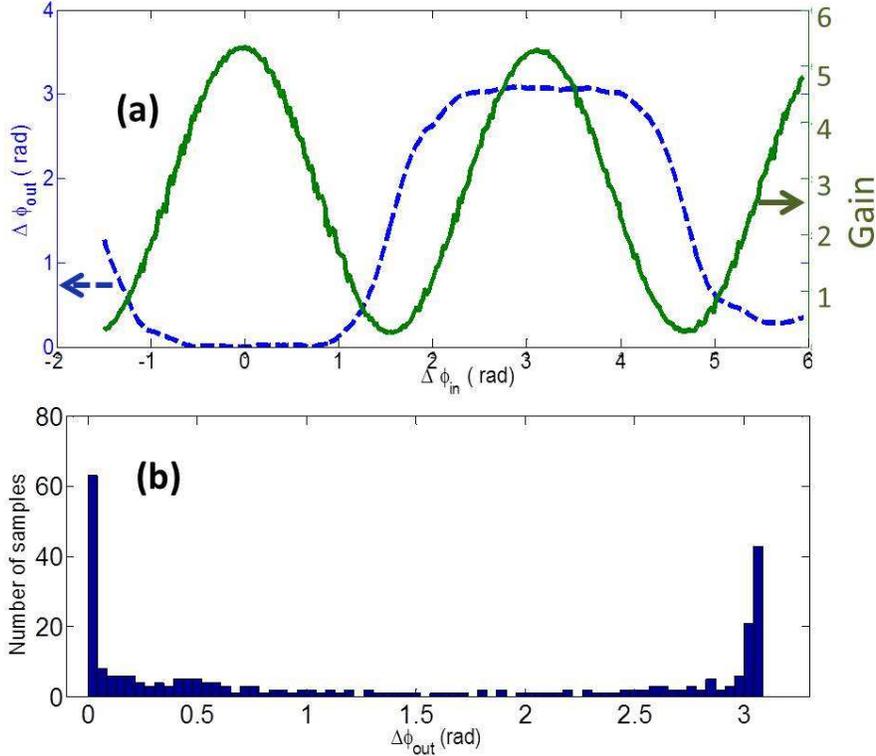}
\caption{(Color online)(a) Variation of the measured output relative phase (blue, dashed) and measured PSA gain (green solid) as a function of the measured input relative phase ($\Delta\phi_{in}$) (b) Output phase histogram}
\label{ph}
\end{figure} 
In order to further explore the quality of this phase sensitive amplifier, we investigate the phase-to-phase transfer characteristics of our PSA. Such measurements have been performed earlier in fiber based PSA in the context of phase regeneration~\cite{PA2011} but for an atom-based PSA, this is being reported for the first time to the best of our knowledge. Such transfer curves can be used to characterize the performance of the amplifier. Figure \ref{ph}a shows the transfer curves (experimental) for a pump power of 30 mW. The blue (dashed) curve shows the phase transfer which is the variation of the output relative phase ($\Delta \phi_{out}$) between the pump and signal with the input relative phase ($\Delta \phi_{in}$), while the green (solid) curve is the corresponding variation of the PSA gain. As we can see, $\Delta \phi_{out}$ is either close to 0 or $\pi$ for a wide range of input phases. For an ideal PSA, when the gain is large, the phase transfer curve is like a square wave, oscillating between 0 and $\pi$, which is called phase squeezing in the telecommunication field~\cite{PA2011}. In terms of a histogram, the output phase is localized around 0 and $\pi$ as shown in fig. \ref{ph}b. Such transfer curves are characteristic of a squeezer~\cite{symplec}: the more localized the output phase, the better the performance of the squeezer.
\begin{figure}[htbp]
\centering
\includegraphics[width=12cm]{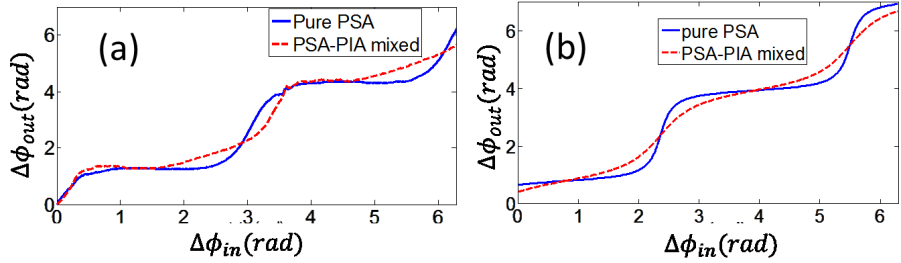}
\caption{(Color online) Phase transfer curves for two different cases , (a) Experimental phase transfer curves for a very good PSA (blue solid) and for PSA-PIA mixed (red dashed) (b) Corresponding theoretical curves}
\label{dpt}
\end{figure}
In fig. \ref{dpt}a, we have plotted such phase-to-phase transfer curves (experimental) for two different cases: in the first case, we send in equal intensities of signal and idler at the input and the measured maximum gain is 5.3. The minimum gain is 0.25, which is close to its ideal value (=0.19). This corresponds to a nearly pure PSA and is represented by the blue solid curve in fig. \ref{dpt}a. The second case (red dashed curve) is a mixed PSA-PIA, as signal and idler intensities at the input are not the same: $I_{s_{in}}/I_{i_{in}}$=1.78. We can see from fig. \ref{dpt}a that for a pure PSA, the output phase is quite flat while for mixed PSA-PIA, the output phase exhibits a higher slope. It should be noted that here for a better physical understanding and a clear comparison, we have not wrapped the output phase like in fig. \ref{ph}a. The corresponding theoretical curves plotted in fig. \ref{dpt}b are obtained considering the fields to be classical ~\cite{PA2011} and agree well with the experimental curves of fig. \ref{dpt}a. The small mismatch of the experimental curve in the case of mixed PSA-PIA with the theoretical curve is probably due to the added uncertainity of the output relative phase between signal-idler which is not completely preserved in the presence of PIA and limits our data processing. Notwithstanding these minor discrepancies, these phase transfer curves qualitatively give an idea of the purity of the PSA. Indeed, we found that these curves are very sensitive to a small mismatch: the output phase is quickly less localized, while $G_{max}.G_{min}$ is still close to 1. From these results, we expect a high degree of quantum squeezing in our output, which will be very interesting for performing quantum information processing tasks using metastable helium. Further, in order to completely model the system, one needs to consider full density matrix for the system and solve Maxwell-Bloch equations: this work is in progress and will be reported later but from our preliminary simulation results, we have found that the Raman coherence does not play much role in the PSA: it is the CPO based processes which mainly contribute to the PSA gain.

In conclusion, we have demonstrated phase-sensitive amplification in metastable helium using ultra-narrow CPO resonance with a maximum gain of nearly 7 and 200 kHz bandwidth, at resonance and within the Doppler width. The measured PSA and PIA gains are well consistent with each other and close to the ideal values, illustrating that we have a pure PSA without any additional unwanted FWM processes. Such large gains in the absence of an external cavity have been made possible due to inherently large CPO-enhanced $\chi^{(3)}$ offered by the system. Further, we have investigated phase-to-phase transfer characteristics which confirm that this system is a very good squeezer. This ensures that we can realize a pure phase-sensitive amplifier which should lead to the generation of non-classical states of light. Since the gain of the PSA is closely related to the degree of squeezing, we believe that we can generate highly squeezed states at low frequencies over some hundreds of kHz. Since optical storage has already been successfully implemented in this system~\cite{MACPO}, it opens the way to realizing an efficient quantum memory~\cite{Lvovsky} using metastable helium with two cascaded ${^4}$He cells. Further, the system is quite versatile and can be used to implement PSA in the degenerate signal-idler configuration, giving rise to the possibility of twin beam generation~\cite{PL2}.

\textbf{Funding} Indo-French CEFIPRA, labex PALM, Délégation Generale à l'Armement (DGA) and the Region Ile-de-France DIM Nano-K, Institut Universitaire de France (IUF), Chinese Scholarship Council (CSC).
\bigskip
% Bibliography
%\bibliography{tri2}

% Full bibliography added automatically for Optics Letters submissions
% Note that this extra page will not count against page length
%\ifthenelse{\equal{\journalref}{ol}}{%
%\clearpage
%\bibliographyfullrefs{tri2}
%}{}

\end{document}